\documentclass[11pt]{article}
 \usepackage{epsf,amsmath,pifont}
  \usepackage[dvips]{graphicx}   

 \usepackage{color}
  
   \newcommand{\BluTn}[1]{\textcolor{blue}{#1}}
    \newcommand{\RedTn}[1]{\textcolor{red}{#1}}

\newcommand{\nn}{\nonumber}                                               
\newcommand{\va}[1]{\langle{#1}\rangle}                                   
\newcommand{\Ds}{\displaystyle}                                           
\newcommand{\gev}[1]{\relax\ifmmode{\text{GeV}^{#1}}                      
                     \else{GeV$^{#1}${ }}\fi}                             

\begin{document}

\begin{center}
{\large \bf Photon Distribution Amplitude in approach with nonlocal condensates
\footnote{
          Talk presented
          at the Workshop on Hadron Structure and QCD: 
          From Low to High Energies (HSQCD 2008), 
          Gatchina, St. Petersburg, Russia, June 30 - July 4 2008. 
          }
}\\[0.5cm]

A.~V.~Pimikov\footnote{E-mail: pimikov@theor.jinr.ru}\\[0.5cm]

{\it Bogolyubov  Lab. of Theoretical Physics,\\
 Joint Institute for Nuclear Research,\\
141980, Moscow Region, Dubna, Russia}\\[0.5cm]

\end{center}
\begin{abstract}
\noindent  We investigate 
magnetic susceptibility
and the leading twist light-cone distribution amplitude
for the real photon in the framework of 
the nonlocal condensate approach at LO accuracy of perturbative expansion.
\end{abstract}
\vspace {1cm}

 \section{Introduction}
   \renewcommand\thefootnote{\arabic{footnote}}
    \setcounter{footnote}{0}
The knowledge of the photon Distribution Amplitude (DA)
      and quark condensate Magnetic Susceptibility (MS)
is quite important for hard exclusive processes involving
 photon emission.
Examples include transition form factors
 like  $\gamma^*\to \pi\gamma$  with one real and one virtual photon ~\cite{KR96,RR96},
  deeply-virtual Compton scattering (DVCS) ~\cite{R96},
   radiative hyperon and meson decays like $\Sigma^+\to p \gamma$, $B\to \rho\gamma$,
    $B\to K^*\gamma$ and $D^*\to D\gamma$~\cite{BFS01,BSBG02,BBK89,R07,R07PRD,BBK88}.

MS $\chi$ has been introduced in the pioneering work~\cite{IS83PLB}:
$ \va{0\!\mid\!\bar q\,\sigma_{\mu\nu}q\!\mid 0}_F =
    \,e_q\, \chi\, \langle \bar q q\rangle \,F_{\alpha\beta}$.
Here $\langle \bar qq\rangle$ is the quark condensate, $F_{\alpha\beta}$ the field-strength
tensor of the external EM field, and the subscripts $F$ indicates that the vacuum expectation value  is
taken in the vacuum in the presence of the field $F_{\alpha\beta}$.
Different theoretical approaches have been used to get this quantity:
QCD SR for proton and neutron magnetic moments~\cite{IS83PLB},
Borel SR analysis of two-points correlator~\cite{BK84, BKY85,BBK02},
the correlator of vector and nonsinglet axial-vector currents~\cite{V03,D05},
instanton liquid model~\cite{GPPRW99} and the instanton inspired model~\cite{D05,GKMS07}.

 Following ~\cite{BBK02,BBK88}, we define the leading twist normalized photon DA $\phi_\gamma(x,\mu^2)$ 
 using the matrix elements of the tensor current with light-like separations:
\begin{eqnarray}\nn
 \,\,\va{0\!\mid\bar q(z)\sigma_{\mu\nu}{\cal C}(z,0)q(0)\!\mid \gamma(q,\lambda)}\Big|_{z^2=0}
 = i\, e_q \chi(q^2) \langle{\bar{q}q}\rangle
      \left(
             \epsilon_{\mu}(q,\lambda)q_{\nu}
            -\epsilon_{\nu}(q,\lambda)q_{\mu}
      \right)
   \!\int^1_0\!\! dx\ e^{ix(zq)} \phi_\gamma(x,\mu^2)\,.
\end{eqnarray}

The Wilson line~\cite{Ste84} ${\cal C}(z,0)=$ 
${\cal P}\exp\left[i g\int_0^z A_\mu(\tau) d\tau^\mu\right]$ is inserted in the matrix element 
for  gauge-invariance.
In the above definitions, $\!\mid\gamma(q,\lambda)\rangle$
is the one-photon state of momentum $p$ ~and polarization vector $\epsilon_{\mu}(q,\lambda)$,
and $\mu^2$ is the normalization point.
The parameter $\chi(q^2 \to 0)$ is
the so-called magnetic susceptibility of the quark condensate.

The photon DA was introduced in~\cite{BBK88},
where closeness of the photon DA to asymptotic form
$\phi_\gamma^{as}(x)=6x(1-x)$ was claimed
on the basis of the standard  QCD SR.
Nevertheless, in instanton model approaches~\cite{ABD06,GPPRW99},
the photon DA has a flat-like form and no end-point suppression.
The authors of ~\cite{BBK02} showed the instability
 of this QCD SR. This instability allows the photon DA to have a non-asymptotic form
 in contrast to~\cite{BBK88}.
Here we develop a nonlocal condensate (NLC) approach \cite{MR86,MR89} to the MS and the 
photon DA 
that significantly improves the properties, and expand the region of applicability of QCD SR.

\section{Nondiagonal correlator in NLC approach}
\begin{figure}[h]
 \centerline{\includegraphics[width=0.47\textwidth]{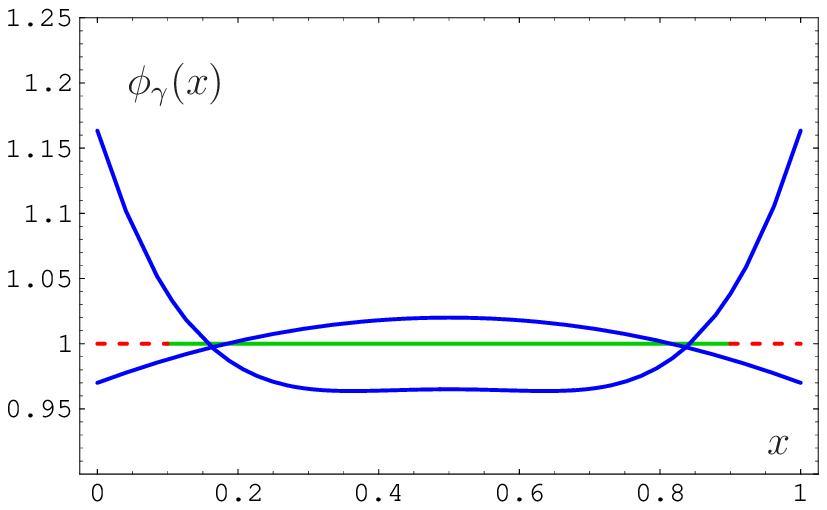}~~~
             \includegraphics[width=0.47\textwidth]{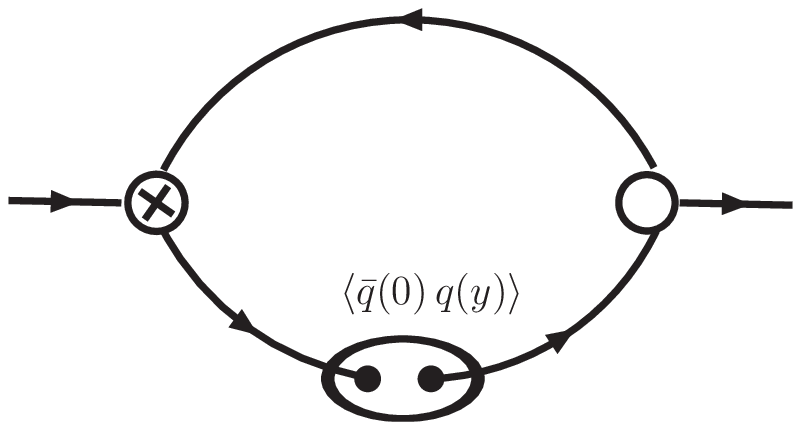}}
   \caption{\footnotesize
           \textbf{Left panel:} Photon DA in different approaches.
               The displayed curves are:
               convex curve  -- instanton inspired model \cite{ABD06},
               concave curve -- instanton liquid model \cite{GPPRW99},
               flat line     -- our LO result in NLC approach.
           \textbf{Right panel:} LO contribution to studied correlator ~(\ref{dif gamma_DA}).
\label{fig:PhotonDAandLOterm}
}
\end{figure}

By using the Background Field formalism~\cite{BBK02} 
one can get an equivalent definition
of the photon DA via a correlator 
of the nonlocal tensor current 
$q(0)\sigma_{\alpha\beta}{\cal C}(0,z)q(z)$ (on the light-cone)
 with the vector one, $j_\mu$,
\begin{eqnarray} \label{dif gamma_DA}
&&\int\!\! d^4y\, e^{iq y} \va{0\!\mid
                       T
                       \left[
                             \bar q(0)\sigma_{\alpha\beta}{\cal C}(0,z)q(z)
                             j_\mu(y) 
                       \right]
                       \!\mid 0}\Big|_{z^2=0}
    =  \\\nonumber
&&~~~~~~~~~~~~~~~~~~~~~~~~~~~~~~~~~~~~~
    i\chi(q^2)\langle{\bar{q}q}\rangle
      \left(q_\alpha g_{\beta\mu}-q_\beta g_{\alpha\mu}
      \right)
   \int^1_0\!\! dx\ e^{ix(zq)} \phi_\gamma(x,\mu^2)\,.
\end{eqnarray}
This nondiagonal correlator can be applied
 for extracting magnetic susceptibility and photon DA
by using the operator product expansion (OPE) method.
A remarkable property of this correlator is 
that the leading-order (LO) contribution is 
completely defined by the nonperturbative vacuum and 
receives no perturbative contributions in the chiral limit at all.
The diagram corresponding to the LO contribution is shown 
in the right-hand side of Fig.~\ref{fig:PhotonDAandLOterm}.
For calculation of this correlator
we would like to apply the NLC technique.
Here we need to introduce only nonlocal scalar condensate:
$
   \langle{\bar{q}(0){\cal C}(0,x)q(x)}\rangle
  \ =\
   \langle{\bar{q}q}\rangle
    \int\limits_0^\infty\!\! f_S(\nu)\,e^{\nu x^2/4}\,d \nu
$
which is parameterized in the general case by 
the distribution functions $f_S(\alpha)$ of vacuum quarks 
in virtualities $\nu$, ~\cite{MR86,MR89}.
Explicit forms of these functions should be taken,
generally speaking, 
from the theory of nonperturbative QCD vacuum
that is still unknown.
In the absence of such a theory we suggested ansatz that
takes into account only the main effect of nonlocality,
the non-zero average virtuality $\lambda_q^2$ of quarks
in the QCD vacuum \cite{BM98}: 
$ f_S(\nu) 
  = 
  \delta\left(\nu-\lambda_q^2/2\right)$.
The nonlocality parameter $\lambda_q^2/2 = \langle{k^2}\rangle=
\int\limits_0^\infty\!\! f_S(\nu)\nu \,d \nu\,$ 
characterizes the average momentum of quarks in the QCD vacuum and
has been estimated in QCD SRs~\cite{BI82,OPiv88} and on the
lattice~\cite{DDM99,BM02}:
$\lambda_q^2 = 0.45\pm 0.05\text{~GeV}^2$.
One can get a relation between the studied correlator at LO (in the r.h.s)
and the magnetic susceptibility multiplied by the photon DA (in the l.h.s):
\begin{eqnarray}
\chi(Q^2)\,\phi_\gamma(x;Q^2,\mu^2)
  = \left\{
\begin{array}{l}
\Ds x \int\limits_0^\infty \frac{d\beta}{\beta}
                                                      f_S\!\left(\beta\right)
                                                      \exp\left(-\bar{x} \frac{Q^2}{\beta}\right)
                               + (x\to \bar{x})\,,  ~\text{ NLC-case} 
\label{eq:NLC-chi}\\
\Ds \frac{\delta(x)+(x\to \bar{x})}{Q^2}, \hspace*{2cm}\text{standard condensates}~(\lambda_q^2\to 0) \label{eq:LC-chi}\\
 \end{array}
 \right. 
\end{eqnarray}
where $\bar{x}\equiv 1-x$, $Q^2=-q^2 >0$.
From (\ref{eq:NLC-chi}) follows that $\chi(Q^2 \to 0)$ has a meaning for the NLC-case
in contrast with the local limit one \cite{BBK02}.

\begin{figure}[h]
 \centerline{\includegraphics[width=0.47\textwidth]{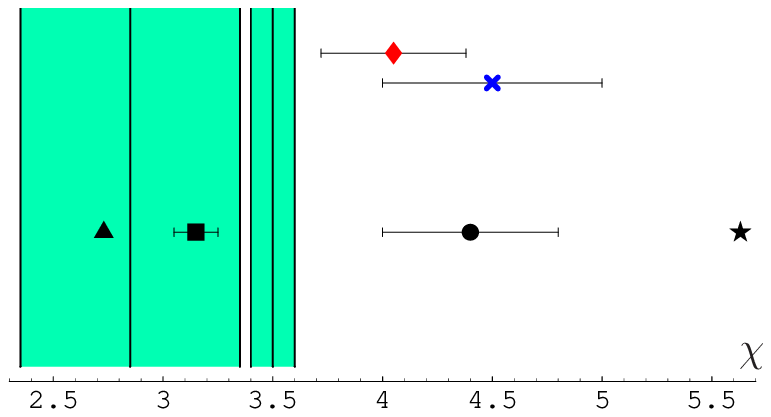}~~~
             \includegraphics[width=0.47\textwidth]{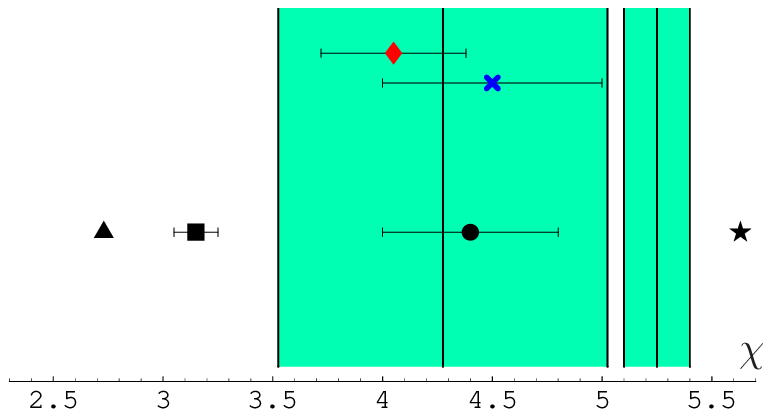}}
   \caption{\footnotesize
            Theoretical constraints~\cite{BBK02,D05,BK84, BKY85,V03} on MS $\chi$
             in comparison with the experimentally based constraints~\cite{R07,R07PRD}. 
             The displayed constraints are:
             \RedTn{\ding{117}} -- our phenomenological estimation (\ref{eq:phen})
              based on $f_\rho$ and $f_\rho^T$ data from ~\cite{BM01},
             \BluTn{\ding{54}}  -- our NLC LO result (\ref{eq:chiLO}),
             \ding{115} -- instanton inspired model~\cite{D05},
             \ding{110} -- Borel QCD SR with local condensate~\cite{BBK02}, 
             \ding{108} -- QCD SR constraints~\cite{BK84, BKY85},
             \ding{72}  -- recent estimation based on 
             AVV-correlator and pion dominance~\cite{V03},
             shaded areas --  experimentally based constraints~\cite{R07,R07PRD}.
           \textbf{Left panel:} Original estimations (shaded area)~\cite{R07,R07PRD} based on asymptotic DA assumption. 
           \textbf{Right panel:} For flat-like DA
            experimentally based constraints (shaded areas) shifted to higher value of MS.
\label{fig:xi}
}
\end{figure}

 To tell the truth,
applying the OPE at low $Q^2\leq \lambda_q^2/2$
can rule out the basis of the approach.
Nevertheless, in this bold extrapolation $Q^2\to~0$ one has
\begin{eqnarray}\label{eq:chiLO}
\chi^{\text{LO}}= \int\limits_0^\infty\!\!d\,\beta\,f_S(\beta)/\beta \stackrel{\rm \delta-\text{ansatz}}{\longrightarrow}
 2/\lambda_q^2=4.5 (0.5)~\gev{-2}
\end{eqnarray}
that provides us a rough but reasonable estimate for $\delta$-ansatz
that agrees with the previous results:
$\chi\approx 2.3-5.6~\gev{-2}$ ~\cite{BBK02,D05,GPPRW99,V03,BK84, BKY85,GKMS07}, see Fig.~\ref{fig:xi} for details. 
 The experimentally based constraints on this value 
$\chi\approx 2.4-3.6~\gev{-2}$ 
was obtained in ~\cite{R07,R07PRD},
where the asymptotic behavior of the photon DA was assumed.
These constraints are shown on the left panel of Fig.~\ref{fig:xi}.

The NLC expressions (\ref{eq:NLC-chi}) can be used to obtain MS and photon DA
at any virtuality $Q^2$ of photon.
As $Q^2\rightarrow 0$ limit 
we get from (\ref{eq:NLC-chi}) a model-independent LO photon DA:
$\phi_\gamma^{\text{LO}}(x)={\theta(1>x>0)}$,
it does not depend on the choice of $f_S(\beta)$ as opposed to MS.
Let us emphasize that this derivation does not fix the $\phi_\gamma^{\text{LO}}$ 
behavior in the vicinity of the end-points $x=0,1$.
This remains unclear due to unreliability of the OPE in this region.
If one uses the $\phi_\gamma(1/2)=1$ assumption,  the experimentally based constraints~\cite{R07,R07PRD} 
on MS will be shifted to higher values $\chi\approx 3.5-5.4~\gev{-2}$,
which is demonstrated on the right panel of Fig.~\ref{fig:xi}.

\section{Magnetic susceptibility from spectral density}

Let us consider an alternative way to get photon properties as an example of MS. 
We write the physical spectral density of this correlation function
as a sum over several narrow resonances plus a smooth continuum starting at a the threshold
$s_0$. Assuming quark-hadron duality, the continuum contribution can be represented by
the perturbative imaginary part of the radiative correction \cite{BBK02}.
As the first approximation one can retain the contribution of two 
lowest-lying $\rho$-meson states only
 and use the same value of the continuum
threshold as obtained in the $\rho$-meson sum rules for the correlation function
of vector currents, $s_0= 2.8\,$GeV$^2$ \cite{BM01}.
Thus, one gets magnetic susceptibility at $Q^2=0$ and $\mu^2=1 \gev{2}$:
\begin{eqnarray}\label{eq:phen}
\chi = -\frac{f_ \rho   f^T_ \rho  }{m_ \rho   \langle{\bar{q}q}\rangle}
       -\frac{f_{\rho'} f^T_{\rho'}}{m_{\rho'} \langle{\bar{q}q}\rangle}
       +\frac{8\alpha_s}{3\pi}\frac{1}{s_0}
       =4.05(33)\,\mbox{GeV}^{-2}\,,
\end{eqnarray}
\begin{eqnarray}
\text{where}~~
\langle{\bar{q}q}\rangle &=& (-0.25\gev{})^3
                              \,,~~~~\alpha_S(\mu=1\gev{})=0.56\,,~~~~~s_0= 2.8\gev{2}\,,\nn\\
   m_ \rho  &&0.7755(4)\gev{}\,,~~~~ f^L_ \rho  = 0.201(5)\gev{-2}\,,~~~~~  f^T_ \rho  = 0.169(5)\gev{-2}\,,\nn\\
  m_{\rho'}&=&1.465(22)\gev{}\,,~~~~ f^L_{\rho'}= 0.175(10)\gev{-2}\,,~~~~   f^T_{\rho'}= 0.140(10)\gev{-2}\,.\nn
\end{eqnarray}
Here we use decay constants that were got in \cite{BM01} by the NLC Sum Rules approach.
The estimate presented in (\ref{eq:phen}) agrees well with the estimate in~(\ref{eq:chiLO}).

\section{Conclusions}
Let us summarize our findings.

(i) 
QCD SR approach with nonlocal condensates at LO accuracy allows 
us to easily get magnetic susceptibility  
as a function of photon virtuality $Q^2$ without singularity 
as $Q^2\to 0$. The corresponding estimate is in good agreement 
with pervious estimations~\cite{BK84,BKY85,BBK02,V03,D05,GPPRW99,R07,R07PRD,GKMS07}.

(ii) 
According to this approach, the photon DA $\phi_\gamma(x)$ 
has a flat-like form at medium $0<x<1$. 
This conclusion agrees with the instanton model methods~\cite{ABD06,GPPRW99}.

(iii) 
This kind of derivation does not fix the $\phi_\gamma^{\text{LO}}$ 
behavior in the vicinity of the end-points $x=0,1$.

\section*{Acknowledgements} 
I would like to thank S.~V.~Mikhailov 
for pointing the problem, 
stimulating discussions and kind support,
 and also A. P. Bakulev, A. E. Dorokhov, and N. G. Stefanis for valuable discussions.
I am indebted to Prof. Klaus Goeke 
for the warm hospitality at Bochum University, where
part of this work was done.
This work was supported in part by 
the RFFR (contract 08-01-00686 and 06-02-16215)
and the DAAD Foundation.

\end{document}